\documentclass[sigconf]{acmart}

\renewcommand\footnotetextcopyrightpermission[1]{} 

\AtBeginDocument{%
  \providecommand\BibTeX{{%
    \normalfont B\kern-0.5em{\scshape i\kern-0.25em b}\kern-0.8em\TeX}}}

\setcopyright{acmcopyright}
\copyrightyear{2021}
\acmYear{2021}

\acmConference[AI in Health '21]{International Workshop on AI in Health: Transferring and Integrating Knowledge for Better Health}{Ljubljana, Slovenia and Virtual}

\usepackage{cleveref}
\usepackage[inline]{enumitem}
\usepackage[htt]{hyphenat}
\begin{document}

\title{Personal Health Data Integration and Intelligence through Semantic Web and Blockchain Technologies}

\author{Oshani Seneviratne}
\email{senevo@rpi.edu}
\orcid{0000-0001-8518-917X}
\affiliation{%
  \institution{Rensselaer Polytechnic Institute}
  \city{Troy}
  \state{New York}
  \country{USA}
  \postcode{12180}
}

\author{Manan Shukla}
\email{shuklm@rpi.edu}
\affiliation{%
  \institution{Rensselaer Polytechnic Institute}
  \city{Troy}
  \state{New York}
  \country{USA}
  \postcode{12180}
}

\author{Jianjing Lin}
\email{linj17@rpi.edu}
\affiliation{
  \institution{Rensselaer Polytechnic Institute}
  \city{Troy}
  \state{New York}
  \country{USA}
  \postcode{12180}
}

\renewcommand{\shortauthors}{Seneviratne, et al.}
\renewcommand{\shorttitle}{Integrating Semantic Web and Blockchain Technologies}

\begin{abstract}
Data integration among various stakeholders in the healthcare space remains a challenge, despite the impressive advances in Health AI in the past decade.
There is a lot of ``messy'' non-standard but structured data that are continually being collected from personal health devices.
While efforts such as the Fast Healthcare Interoperability of Resources (FHIR) are underway in standardizing the data representation formats, there is currently a gap in the standard in addressing the health data ecosystem's decentralized nature. 
As we see explosive growth in chronic diseases such as diabetes, healthcare providers need Observations of Daily Living (ODL) of their patients to treat them effectively.
The best way to obtain ODL is through personal health devices.
However, such devices are manufactured by various device makers, and they may not follow standards or integrate with existing Electronic Health Record (EHR) systems.
It is also imperative that any data sharing that happens will occur in a secure and trustworthy environment, without being too restrictive, i.e., tied to a particular EHR vendor.
This paper presents a scalable solution to bridge this gap using a system that implements semantic web and blockchain technologies.
Our solution uses FHIR compliant semantic web based data templates in conjunction with smart contracts on the blockchain to provide healthcare providers with insights on their patients' daily activity that cannot be readily determined solely through patient encounters at the clinic.
\end{abstract}


\maketitle

\section{Introduction}

Health information is increasingly available through EHRs, with a growing number of EHR systems connected through health information exchange networks currently dominated by commercial vendors.
One of the barriers to health record exchange is the need for shared data standards to enable systems to exchange documents and correctly interpret them.
Fortunately, there is a transition toward shared interface standards to enable exchange. This design methodology increases usability by defining the document structure returned from a query without dictating the underlying system design. Shared Application Programming Interfaces (APIs) enable data retrieval regardless of participating systems' design. At the forefront of this change is FHIR, a Health Level-7 (HL7) project defining a set of data resources and APIs for accessing them. 
FHIR improves flexibility over previous exchange protocols by introducing a higher level of abstraction, allowing data to be represented in a unified format independent of the underlying data architecture used by the health system implementing the API.
FHIR also supports representation formats that are friendly to semantic web applications.

Further increasing the quantity of health information per patient is led through medical devices, which can wirelessly transmit sensor data through the ``The Internet of Things'' (IoT) guidelines. One of the main advantages of such devices is that they facilitate innovative AI, machine learning, mathematical and statistical approaches to building predictive patient physiological learning models, which improve health outcomes.
Integrative technology can liberate healthcare providers from the growing burden of clerical work and synthesize patient data, including behavioral, genomic, microbiomic, and so on, into genuinely personalized healthcare.
On the patient's end, healthcare devices such as insulin pumps can be wirelessly connected to glucose meters to automatically inject insulin into diabetic patients to improve their quality of life without the burden of manually checking the readings and administering the medication themselves.
Most importantly, however, the healthcare provider can use this information to provide better health outcomes for their patients because the only quantifiable, reliable, and accurate data a they currently have are lab results and measurements taken in the office. Medical device data have the potential to provide more granular and relevant day-to-day information to the healthcare provider, which can be used to make better patient decisions. 

However, this ideal future is hindered by a lack of integration between medical devices and other electronic medical record systems. We have not found any personal medical devices currently integrated with EHR systems during our search for possible integration solutions present in the market. This is true not only for smaller device manufacturers, but is similarly true for larger companies such as Medtronic and Epic systems~\cite{20_2017}. This lack of integration prevents vital patient information from reaching the healthcare provider, preventing the provider from making the best health decision possible. As we further elaborate in later aspects of this paper, we believe that there is unrealized potential for personal medical devices to be adopted throughout the healthcare industry. 
Furthermore, as we have witnessed from the COVID-19 pandemic, there is an increasing need to have reliable at-home monitoring because coming to a clinic may pose a grave risk to specific at-risk populations.
Therefore, there is a significant need to have reliable data streams from various devices integrated to provide a cohesive picture of the patients. To solve this problem, we have developed a prototype system called \emph{BlockIoT} that can merge these various health data streams, store patient data securely on a decentralized storage system, and represent patient analytics in an EHR system through the use of smart contracts. BlockIoT can serve as a bridge between the data produced by medical devices, and the healthcare providers who view this data through FHIR-compliant EHR systems.



\section{State of the Art}
\label{sec:soa}

As our focus is to develop a system that brings the best of both semantic web and blockchain technologies, we have investigated state of art in each field and prior work that attempts to combine the two technologies.

\subsection{Semantic Web for Healthcare}
\label{sec:sw}

The semantic web vision entails a logic-oriented framework for representing and connecting heterogeneous knowledge resources using explicit semantics, simplified annotation, sharing of findings, rich explicit models for data representation, aggregation and search, and easier reuse of data in unanticipated ways~\cite{berners2001semantic}. 
Ontologies in healthcare and semantic web technologies, in general, are used to share, integrate and reuse biomedical data to enhance diagnostic procedures, reduce costs, and enhance research in the biomedical field.
One of semantic web technology's best use cases is aggregating information from multiple sources to provide a cohesive summary of relevant clinical details for a patient. It has been explored in the synthesis of heterogeneous operational and medical information and knowledge resources to orchestrate patient-specific healthcare plans~\cite{abidi2006adaptable}.
Given that the healthcare knowledge requisite for effectively treating a patient exists in multiple places in various modalities, semantic web technology is an ideal solution to integrate all of this data and build a cohesive picture of a patient's health.
To support such data integration in the clinical application space, Systemized Nomenclature of Medicine - Clinical Terms (SNOMED-CT) aims to support clinical information recording using a controlled vocabulary that enables machine interpretation for information exchange, decision support, aggregation, and analysis~\cite{donnelly2006snomed}.
However, in practical clinical applications, delivering extensive terminology such as SNOMED-CT has presented several issues, including challenges with term search, the general preference for more ubiquitous relational database systems, and the performance bottlenecks in description logic based reasoning systems~\cite{wroe2006semantic}.
In addition to SNOMED-CT, several other ontologies and terminologies have been introduced in healthcare for describing and integrating medical data, such as Unified Medical Lexicon System (UMLS)~\cite{bodenreider2004unified}, Foundational Model of Anatomy (FMA)~\cite{rosse2003reference}, and International Classification of Diseases (ICD)-11~\cite{tudorache2013using}. 
Standards such as OpenEHR~\cite{kalra2005openehr} have integrated the main biomedical ontologies.

Several specialized applications combine semantic web technologies with wearable devices, and some examples are outlined below.
A prototype clinical decision system that estimates cardiovascular disease risk by integrating recommendations to patients using a reasoning engine based on Web Ontology Language (OWL) and Semantic Web Rule Language (SWRL) using a mobile phone and a Bluetooth sensor to monitor blood pressure is described in~\cite{hervas2013mobile}. 
An ontology-based ambient intelligence framework supporting real-time remote monitoring of patients diagnosed with congestive heart failure, which provides personalized medication and an intelligent emergency alerting of the dedicated healthcare provider, is described in ~\cite{hristoskova2014ontology}. 
An architecture for monitoring patients at home using an ontology that unifies the management procedure to integrate incoming data from various sources and a communication layer to access and exchange information modeled is described in~\cite{lasierra2013designing}.
A portable data provenance toolkit supporting provenance within health information exchange systems is proposed in~\cite{tyndall2018fhir}, which claims to offer a solution to address shortcomings of past efforts, including mapping complex datasets and enabling interoperability via an exchange.
The ``HealthIoT Ontology'' represents medical device data related to a person's health~\cite{rhayem2017healthiot,rhayem2017ontology}. This ontology reuses concepts from the Semantic Sensor Network Ontology~\footnote{\url{https://www.w3.org/TR/vocab-ssn}} to expand specifically to the health domain along with rules that define conditions for medical alerts in an ``IoT Medicare'' clinical decision support system. However, the ontology does not conform to the FHIR standard~\cite{bender2013hl7}, which will likely be the data representation format adopted more broadly in the future.

Given the above analysis, we assert that the widespread adoption of semantic web technologies in healthcare applications, particularly healthcare that leverages personal health devices, is yet to be fully realized, thus provide ample space for innovation as proposed in this paper. 

\subsection{Blockchain for Healthcare}
\label{sec:blockchain}

Blockchain technology has been a technological disrupter in the past decade. Its most successful use case has been to transfer currency between two mutually distrusting parties, with the bitcoin cryptocurrency dominating the space.
As bitcoin's popularity led to more features that inspired the development of programs that govern the rules of transactions on the platform, including a non-Turing complete scripting language, which is quite limited, as it only features some basic arithmetic, logical, and crypto operations (e.g., hashing and verification of digital signatures)~\cite{bartoletti2017analysis}. 
Therefore, more expressive logic was needed for complex applications such as healthcare applications.
Smart Contracts, i.e., programs whose correct execution is automatically enforced without relying on a trusted authority, are being used to address this challenge~\cite{buterin2013ethereum}.
The features available in modern smart contract languages, such as Solidity, are necessary for purely data-driven transactions, such as the applications we need in healthcare.
However, the holy grail of blockchain healthcare applications appears to be for electronic medical records, cross-system interoperability, patient-centric control of records~\cite{mcghin2019blockchain}, and clinical trial management~\cite{wong2019prototype}. 
Purchasing medical devices in the pharma supply chain~\cite{bocek2017blockchains} and the internet of medical things have shown much value and are quickly gaining traction.
One early pilot by Napier Edinburgh Napier University, Spiritus Partners, and the National Health System (NHS) in Scotland shows blockchain's value to medical devices~\cite{clauson2018leveraging}. 
The blockchain pilot enabled permissioned users at each hospital to see in real-time where each device was located and which staff were trained on them for use.
Another example of a blockchain application is the Gem Health Network that used the Ethereum Blockchain Technology to create a shared network infrastructure~\cite{mettler2016blockchain}. The Gem Health Network allows healthcare professionals to access the same healthcare information removing the limitations of centralized storage using the blockchain-based \emph{GemOS}~\cite{shi2020applications}.
Similarly, several conceptual efforts and smaller real-world products and pilots are working to bring blockchain technology to healthcare, such as OmniPHR~\cite{roehrs2017omniphr}.

A blockchain-based prototype system that illustrates how patients can manage their healthcare records that are source-verifiable is discussed in the medical record management system proposed in~\cite{kuo2017blockchain}.
A ``Healthcare Data Gateway,'' which would enable patients to manage their health data stored on a private blockchain, is described in ~\cite{yue2016healthcare}.
Healthcare Data Gateway (HGD) facilitates patients to take control of their medical records securely~\cite{yue2016healthcare}, where a smartphone application uses a unified non semantic web based data schema to help store and organize data in a database management system. 
A comparable system that uses a public blockchain implementation, where healthcare data is encrypted but stored publicly, creating a blockchain-based \emph{Personal Health Record} is described in~\cite{ivan2016moving}.
MedChain is another example where a permissioned network of stakeholders (including the patient) is used to facilitate medication-specific data sharing between patients, hospitals, and pharmacies~\cite{gordon2016secure}. 
Other systems such as MedRec \cite{azaria2016medrec}, too, enables patient data sharing and incentives for medical researchers to sustain the system using a blockchain. 
MedRec incentivizes medical stakeholders to participate as miners by providing them access to aggregated anonymized data as mining rewards in return for sustaining and securing the network.
The \emph{Pervasive Social Network} is a system developed for healthcare settings comprising mobile computing and wireless sensing that establishes data sharing among smart devices and sensors between a patient and their healthcare provider with the help of a blockchain~\cite{zhang2016secure}.
MeDShare, a system designed to address the issue of trustworthy medical data sharing among medical professionals that store data, focuses on smart contracts to determine and monitor access to the patient records in prototype EHR systems~\cite{xia2017medshare}.
A permissioned blockchain focused on oncologic care that leverages off-chain cloud storage for clinical data, using the blockchain to manage consent and authorization, is proposed in~\cite{dubovitskaya2017secure}.
A blockchain-based distributed electronic medical records searchable scheme is proposed where access to the encrypted health records is enabled through a keyword search~\cite{sun2020blockchain}. The healthcare provider is responsible for generating medical information for the patient and extracting keywords for the patient’s medical information after obtaining the patient’s consent. 

As evidenced by the above list of blockchain-based healthcare solutions above, there is a lot of enthusiasm in this space.
However, none of these solutions have been developed with interoperability in mind. The decentralized architectures proposed in each system appear to be fragmented when taken as a whole.
Furthermore, there are many benefits if the data pipelines are streamlined with standards-based, preferably semantically-annotated, ontologies to elicit health data intelligence automated through smart contracts.

\subsection{Semantic Web and Blockchain for Healthcare}
\label{sec:swandblockchain}

One notable example of an attempt to combine some aspects of semantic web and blockchain technologies is the work by Peterson et al., who have examined interoperability for the fruitful exchange of healthcare data within the structure and semantics of the data transferred~\cite{peterson2016blockchain}. They have devised a blockchain infrastructure catered specifically to the exchange of healthcare data that utilizes a consensus mechanism called \emph{Proof of Interoperability}. While this consensus mechanism appears to be very useful, practical implementation of it appears to be challenging.
Another example is the FHIRchain, which presents an Ethereum based solution designed to meet the Office of the National Coordinator for Health Information Technology's requirements by enforcing the use of FHIR to share clinical data by validating whether the generated reference pointers to sensitive patient records that are stored off-chain follow the FHIR API standards~\cite{zhang2018fhirchain}. 
We note that even though the FHIR has semantic web serializations, FHIRchain does not consider the semantics of the FHIR Resource Description Framework (RDF) vocabulary, as is the aim of the work presented in this paper. 

Apart from the above examples, we did not come across many research and development efforts that combine some aspects of semantic web and blockchain technologies as it was the case for each technology described in \Cref{sec:sw} and \Cref{sec:blockchain}.
However, we believe that combining the two technologies provides a potent concoction that will provide immense benefits to the healthcare industry and is ripe for investigation.
Therefore, the work presented in this paper is one of the first efforts at cohesively combining the two fields, thus setting our work apart from all the existing implementations and research.

\section{System Design}

\begin{figure*}[!tbph]
    \centering
    \includegraphics[width=\textwidth]{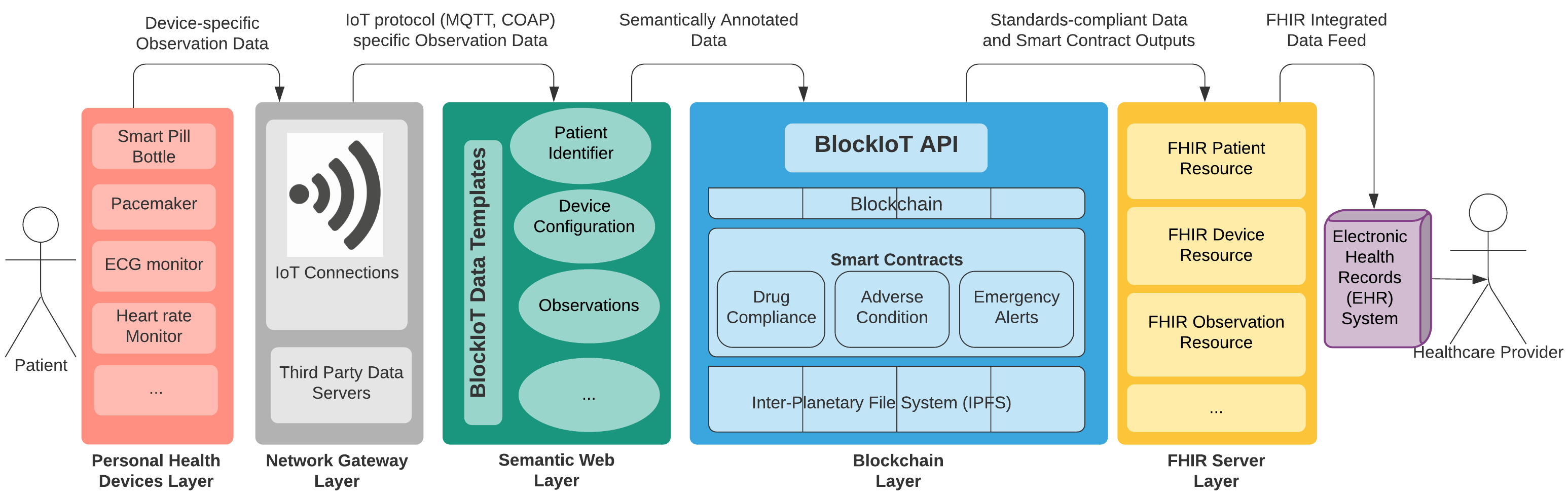}
    \caption{System Architecture}
    \label{fig:blockiot-architecture}
\end{figure*}

We present an innovative information transfer system, \textbf{BlockIoT}, that combines semantic web and blockchain-based technology to enable connectivity between personal health devices at the patient's end and the EHR system at the healthcare provider's end. 
BlockIoT aims to transform the medical field from an episodic, reactive, disease-focused, fragmented, and hospital-centered field to a proactive, preventive focus on well-being and quality of life, patient-centered and interoperable field. 
BlockIoT contains several layers as depicted in \Cref{fig:blockiot-architecture} designed with the following guiding principles in mind.

\begin{itemize}
    \item \textbf{Trust:} allow only designated people or services to have device or data access 
    \item \textbf{Identity:} validate the identity of people, services, and ``things''
    \item \textbf{Privacy:} ensure device, personal, and sensitive data are kept private
    \item \textbf{Protection:} protect devices and users from physical, financial, and reputational harm
    \item \textbf{Safety:} provide safety for devices, infrastructure, and people
    \item \textbf{Security:} maintain security of data, devices, institutions, systems, and people
\end{itemize}

The layered architecture of the BlockIoT system provides several benefits when scaling the system in an interoperable way as we consider new personal health devices to support specific healthcare use cases.
It further provides the necessary semantic abstraction when representing the data gathered at each layer cognizant of the standards-based or proprietary systems that may be used either at the patient's end or the healthcare provider's end.
We discuss the functionality of each of the layers below.

\subsection{Personal Health Devices Layer}

Mobile Health (mHealth) is a growing field in healthcare applications involving devices such as miniaturized sensors, low-power body-area wireless networks, and pervasive smartphones that serves to collect and transmit patient health data from the patient's location (such as the patient's home) to a specific destination (such as commercial servers of the medical device company) using IoT. 
For example, a medical device can be a wearable heart-rate monitor that can transmit real-time heart rate to a specific destination (identified in the \textbf{Network Gateway Layer}) over WiFi. 
However, mHealth suffers from many of the problems that the broad healthcare centralized server systems suffer. The specific problems are data sharing and consent management, access control, authentication, and user trust.
The \textbf{Personal Health Devices Layer} handles these aspects by handing off device-specific observation data to the \textbf{Network Gateway Layer} in a reliable way.

Consider a patient with multiple co-morbidities, such as diabetes, hypertension, chronic obstructive pulmonary disease (COPD), and heart diseases. To manage their condition effectively, they may use a multitude of devices. The following are some of the personal health devices that may be at our patient's disposal for each disease. In this example, each device has the ability to transmit data points over a connection such as WiFi. 
\begin{itemize}
    \item \textbf{Diabetes:} Glucose Meter, Continuous Glucose Monitor, Insulin Pump, Weight Scale
    \item \textbf{Hypertension:} Blood Pressure Monitor, Weight Scale, Heart Rate Monitor 
    \item \textbf{Heart Failure:} Blood Pressure Monitor, ECG, Heart rate monitor 
    \item \textbf{COPD:} Spirometer, Peak Flow, Thermometer, Pulse Oximeter, Heart rate monitor, CO2 monitor
\end{itemize}
As shown in the above list, there are several overlaps in the devices used in monitoring the conditions pertaining to each of the diseases.
It is imperative to capture all such readings and provide targeted alerts based on a specific disease's criteria.
This layer facilitates the data transfer coordination from these various devices by integrating with the device APIs and converting the observation data collected into a semantic web format in subsequent layers. 

\subsection{Network Gateway Layer}

This layer is tightly coupled with the \textbf{Personal Health Devices Layer} and acts as an interface for all the receiving endpoints of all incoming medical device data. Because of the immense number of medical devices available today, the API was created with flexibility in mind. As a result, multiple IoT communication protocols such as Hypertext Transfer Protocol Secure (HTTPS), Message Queuing Telemetry Transport (MQTT)~\cite{katsikeas2017lightweight}, and Constrained Application Protocol (CoAP)~\cite{bormann2012coap} are implemented in this layer to ensure maximum possible coverage of most medical devices present in the market.

\subsection{Semantic Web Layer}

The Semantic Web Layer consists of two parts: 
\textbf{Configuration files} and \textbf{Templates}. 

\emph{Configuration files} are simple key-value JavaScript Object Notation (JSON) files that can be thought of as ``translators'' that convert proprietary medical device syntax into system-understandable data. These files are essential to maintain the flexibility of the system. 
Each medical device/server would also need to be authenticated with BlockIoT's protocols. 
\emph{Templates} are generated for each medical device to translate external medical device syntax into properly defined terminology. This terminology can then be converted as protocols such as FHIR~\cite{bender2013hl7} and SNOMED-CT~\cite{donnelly2006snomed}, and can be exported to EHR systems (this process is further discussed in \Cref{sec:fhir-server}.

\noindent First, each configuration file consists of:

\noindent\textbf{Identifiers} that are based on IEEE Standard 11073-10101-2019~\cite{IEEE11073}. Each medical device must have some form of patient identification number and a medical device identification number, which is used to determine which patient uses which device. Typically, this is already set by the medical device manufacturer, who may use it for tracking purposes. However, by taking advantage of these identifiers, one can identify the patient without requiring the manufacturer to assign another identification number for the system.

\noindent\textbf{Device Configuration} contains a key representing unique identifiers in the device data and other device-specific parameters such as the specialization (blood pressure), manufacturer, model, serial number, firmware, hardware, software, time properties (clock type, synchronization, resolution, accuracy), and regulatory information. Each type of device will have its separate template. For example, a heart rate sensor will have a template that will contain all types of information transmitted by the heart rate monitor (such as beats per minute and SpO2).

\noindent\textbf{Key-Value Parameters} translate the medical device syntax into a specific set of defined terminologies. For example, the five parts to an electrocardiogram (ECG) such as \texttt{p},\texttt{q},\texttt{r},\texttt{s}, and \texttt{t}, can be defined in different languages (binary or hexadecimal) and syntactically many different ways (for a computer, \texttt{p} and \texttt{p-wave} are fundamentally different). As a result, it is crucial to define what exactly each character in the medical device's output represents.

\noindent Second, the \textbf{Observations} template contains the patient's body measurements and manual, self-reported vital signs. These observations may be in the form of episodic (typically up to a few times per day) scalars or short-vector vital signs measurements, notably weight, blood glucose level, blood-pressure plus pulse, and ECG rhythm strip. It may also contain automatic, device-reported vital signs obtained from wearable health devices. This template is also used to identify specific physiological parameters or guidelines. These guidelines could represent which action should be taken if specific medical parameters are not within the normal limit. For example, if a patient's blood pressure significantly increases than what is considered ``normal,'' an emergency alert will be sent to the provider regarding the patient's blood pressure (this is further discussed in the \textbf{Blockchain Layer}). Because of a patient's possibly complex physiology, parameters can be set through machine learning algorithms based on multiple variables such as gender, age, height/weight, previous health exacerbation (such as a stroke), and past health-related conditions of the patient. A template is generated for each possible physiological measurement, such as:
\begin{enumerate}[label=(\roman*)]
  \item Scalar (e.g., \texttt{37.5 C, 150 lbs, 96 mg/dL, 10 km, 10943 steps, 53 bpm})
  \item Vector (e.g., blood pressure measurements such as \texttt{\{102 mmHg, 51 mmHg, 76 mmHg\}} )
  \item Code (e.g., \texttt{PREPRANDIAL, FASTING, CASUAL, BEDTIME})
  \item Event/State (e.g., \texttt{battery\_low}, \texttt{poor\_perfusion}, \texttt{signal\_poor}, \texttt{pulse\_irregular})
  \item Waveform (e.g., ECG observation)
  \item String (e.g., \texttt{Fat Burning, Hill Climb})
\end{enumerate}

\subsection{Blockchain Layer}

While the data provided through the semantic web layer is highly integrative, unless the ownership of the data is verified, it may not be trustworthy as asserted in~\cite{eysenbach2003semantic}.
Therefore, we use blockchain technologies, namely distributed storage~\cite{benet2014ipfs} and smart contracts~\cite{buterin2013ethereum} via the BlockIoT API, to achieve the end-to-end trustworthy sharing of personal health data from the patient to the provider. 
We build our prototype on the Ethereum blockchain, which currently uses the Proof-of-Work (PoW) consensus mechanism~\cite{ethereum-pow}.

\noindent\textbf{BlockIoT API}
The BlockIoT system is constructed as a combination of an API and various smart contracts that connect existing EHR systems to several medical devices. 
The API facilitates batch delivery of measurements to the smart contracts deployed on the chain and for use by the healthcare provider, and eventually without measurements getting lost or corrupted.
It is primarily characterized as a decentralized API to facilitate the coordination of endpoints that send and receive data published by a patient's medical IoT device. 

\noindent\textbf{Decentralized Storage} is the process of breaking up the storage of records from a central server to multiple servers through blockchain's ledger, and can facilitate faster access to medical data, system interoperability, patient agency, and improved data quality.
We implemented the storage for BlockIoT using the Inter-Planetary File System (IPFS)~\cite{benet2014ipfs}.
The combination of IPFS and blockchain allows users to process large amounts of data through IPFS so that the data itself does not need to be placed on the chain, saving the network bandwidth of the blockchain and effectively protects it.
The IPFS network is a fine-grained, distributed, and easily federated content distribution network common to all data types and has fewer storage restrictions.
IPFS  was the primary choice of the BlockIoT system due to all such features and its ease of use and ability to make data resilient over time. However, because the IPFS hashes change as the content of the file changes, Inter-Planetary Name System (IPNS)~\cite{ipns} is used to ensure that the address of patient content is the same regardless of the patient data change.
In our system, each patient is connected through a peer-id in IPNS generated based on their biometrics (first name, last name, and date of birth).
The patient's folder containing the personal health data transfer transactions to the EHR (not the actual data itself) is stored in IPFS and published with the IPNS key.

\begin{figure}[!tbph]
    \centering
    \includegraphics[width=\columnwidth]{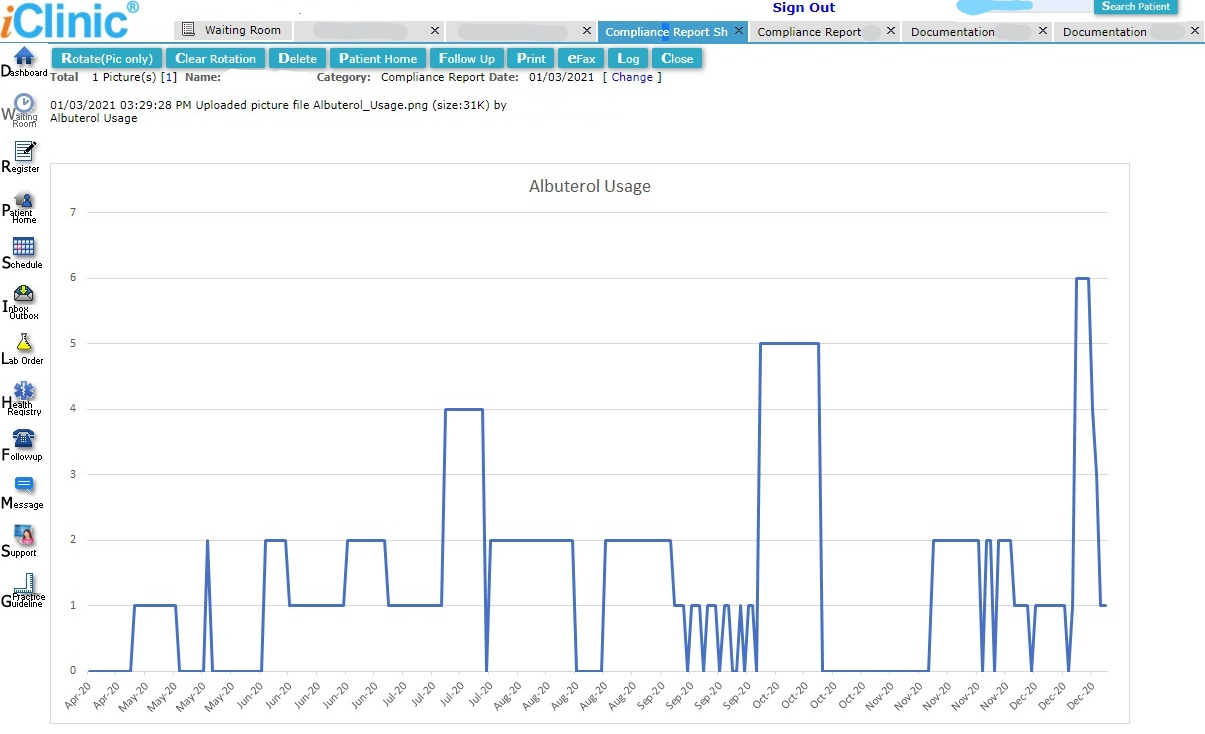}
    \caption{Drug Compliance Chart in a Prototype EHR System Integrated with the BlockIoT API}
    \label{fig:drug_compliance}
\end{figure}

\noindent\textbf{Smart Contracts} provide three main functions in the system. 
First, they provide the access control mechanism that enables the personal health data transfer between the patient and the health care provider. All the parties sending, and accessing the data, must be authenticated via a decentralized application that implements the BlockIoT API.
Second, BlockIoT implements several useful alerts based on the data transmitted from the patient's end.
Unlike point-of-care devices available at the clinic that are already integrated into the EHR systems, such standard EHR systems may not have the requisite functionality to interpret these patient-centric values obtained from various devices. Therefore, smart contracts can bridge this gap and provide alerts to the healthcare provider or any other authenticated and authorized stakeholder. 
For example, if the patient uses a smart pill bottle, the \textbf{\emph{Drug Compliance}} smart contract can determine any deviations from the adherence guidelines (which checks for whether the patient took their medication). These guidelines will be encoded in the ``compliance'' template and be composed of parameters that contain normal limit numbers to determine whether a data point is considered ``normal'' or outside of the expected values.
If the patient uses a heart rate monitor, the \textbf{\emph{Adverse Condition}} smart contract can detect an abnormal heart rhythm.
If the patient is elderly and wears a fall detection sensor, the \textbf{\emph{Emergency Alert}} smart contract will autonomously spring into action alerting someone who may help the patient in a time-sensitive manner.
Third, while it is possible to send over raw medical device data to an EHR system, the data sent to the EHR system are not useful to the healthcare provider as is. It is better to facilitate the provider by providing them with vital statistics or charts. These summarized outputs enabled through smart contracts can quickly allow a provider to review the patient's health metrics and decide further treatment as necessary.
In all these instances, the smart contracts can be customized with the required parameters with the help of the semantics-based BlockIoT templates, but more importantly, they can be programmed to take the necessary actions, such as alerting emergency services, the patient's healthcare provider, or a loved one. 
Healthcare provider accessible graphs are generated and are exported to the EHR system to facilitate this process, based on templates created for each type of data as depicted in \Cref{fig:drug_compliance}. 

\begin{figure}[!tbph]
    \centering
    \includegraphics[width=\columnwidth]{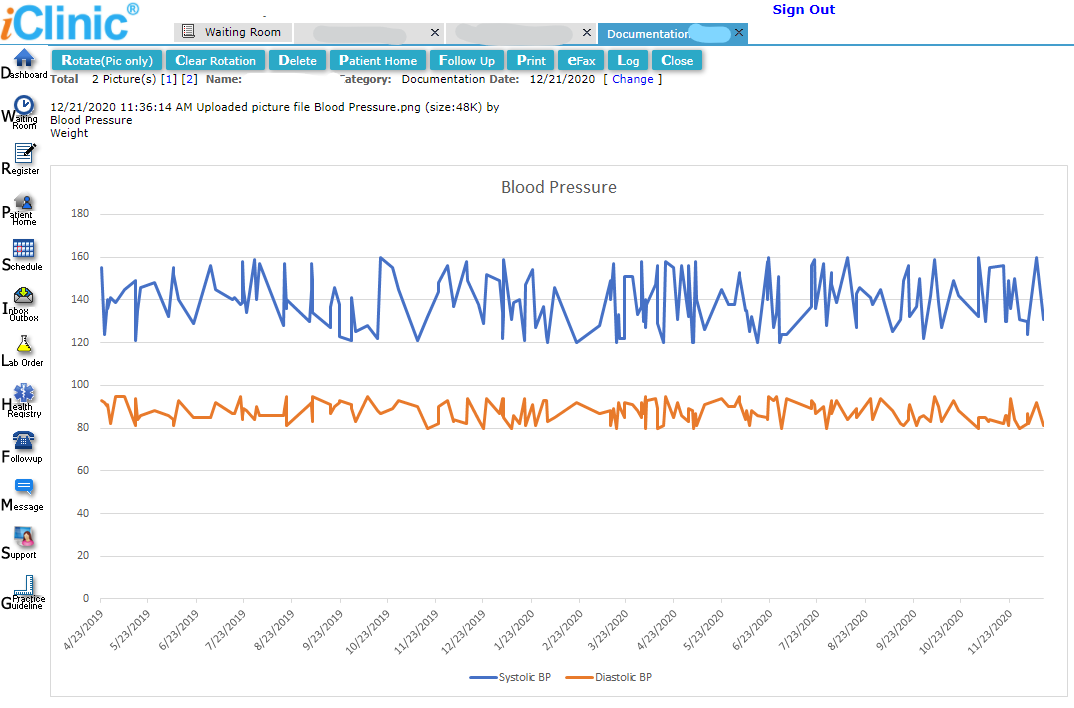}
    \caption{Blood Pressure Chart in a Prototype EHR System Integrated with the BlockIoT API}
    \label{fig:bp}
\end{figure}

\subsection{FHIR Server Layer}
\label{sec:fhir-server}
The basic building block of FHIR is a \emph{resource}. 
There are approximately 100 FHIR resources, and while they can be used for various purposes, they are typically used to exchange clinical content such as encounters, care plans, and diagnostic orders. 
FHIR's query specification enables intuitive data retrieval and automatic grouping of related resources into a single response document.
Point-of-care devices from hospitals, patient care devices (e.g., thermometer, weight scale, pulse oximeter or ECG monitor), and personal health devices (e.g., fitness trackers, home digital blood pressure cuffs) are all a part of the \emph{Devices on FHIR effort}.
The FHIR standard is currently undergoing testing to connect medical devices and EHR systems as well~\cite{ihe-usa}. 
FHIR is a powerful asset for this layer because its rich data set enables reliable record delivery, while abstracted data mapping allows health information systems to use their choice of underlying databases. 
FHIR as a data model for other personal health use cases in the remote patient monitoring domain, e.g., the patient-reported outcomes measures, is gaining traction~\cite{pchalliance} but there is a gap that needs to be filled in securely and efficiently connecting personal health devices with EHR systems.
In the BlockIoT implementation, the chosen data storage mechanism is distributed storage through IPFS that addresses some of these challenges. 

An EHR system is defined as a centralized system whose purpose is to store and integrate electronic patient health records and represent them to a healthcare provider. Here, an EHR system would receive data from the BlockIoT system through smart contracts that control the EHR's access to data via the FHIR Server Layer. 
We have used the FHIR semantics in the intermediate layers to successfully map the data transformations into a format that the EHR systems can understand as long as they implement the FHIR specification.
As many EHR systems implement FHIR soon, the BlockIoT API will be used to convert all the data transferred between the various layers into FHIR format, ready to be consumed by the healthcare provider's EHR system.

\section{Discussion}

Health information is collected by various people in a variety of locations in all sorts of different ways. 
It is often complicated, messy, and unorganized. 
The messiness is sometimes not apparent when looking at data from hundreds of thousands of individuals in aggregate. 
Unless someone is familiar with the data, they may have questions about where the data is coming from, what data can be coupled, merged, combined, and what caveats.
These aspects should be encoded in some form of metadata scheme, preferably using semantic web technologies.
Furthermore, any volume of data can be crunched into statistical AI programs for some insights, but these insights may not be explainable due to the lack of inherent semantics. 

For example, consider a simple bit of health information, such as a reading of systolic over diastolic blood pressure: 120/80. 
This blood pressure can be collected in various states of the patient's life: at home with their smartwatch's blood pressure monitor after 30 minutes of relaxing, at the machine in the pharmacy after 30 minutes in traffic, at the doctor’s office after a morning fast, at a minute clinic after a burger and fries, at the hospital while under anesthesia for surgery, or in the park via Fitbit during a long run.
This data represents the same kind of information, but represents different states of a person’s physiology, recorded by different devices of different validity and accuracy, under different supervision types from self, reported to readings captured by a medical professional. 
The data represents different things and therefore have different implications. For example, the interpretation of a blood sugar level in the morning before a meal, and a reading right after a meal is significantly different. 

Unfortunately, this data in the current state of the art is stored in different data silos in different formats, making it next to impossible for data merging to get valuable insights about the person's physiology.
Therefore, as such supplementary health information is becoming more easily obtainable through smart devices, and patients are traveling to multiple healthcare providers, the sharing and privacy of patients' health information need a robust solution such as the BlockIoT system proposed in this paper. 
Given the data harmonization enabled through the BlockIoT data templates, a healthcare provider may be able to see the longitudinal view of the patient's data collected through various such devices as depicted in \Cref{fig:bp}.

We have addressed interoperability, data sharing, the transfer of medical records within various devices and disease contexts to make the data transfer process more streamlined, secure, and transparent between all the stakeholders involved.
Interoperability is the process of sharing and transferring data among different sources, and it is an essential requirement for the healthcare industry that has so far been unrealized to its full potential.
Patients do not have a unified view of these scattered records, and this also applies to healthcare providers, as they do not have access to up-to-date data regarding the patients if the records are located elsewhere.
Our development focus was towards shared datasets and architecture-independent interfaces, including blockchain. 
This methodology increases usability by specifying shared APIs to enable structured queries regardless of the participating systems (i.e., wearable personal health devices and EHRs)
Adopting a RESTful API query structure facilitated by the BlockIoT API simplifies queries, facilitates faster integration, and reduces the number of queries required to retrieve a relevant collection of resources. 
An architecture-independent query API will enable interoperability between systems regardless of how they store and structure their internal data.

There is immense value in integrating data collected from medical devices, as these devices represent islands of information containing hundreds of parameters.
These wearable devices typically use IoT features because the data from the devices are stored in the cloud.
Ideally, these devices' utility should span from the home to the hospital that reports relevant physiological, environmental, and patient-generated parameters.
Much of the data integration that has happened so far offer expensive, complex, and brittle integrations.
The devices considered for BlockIoT data template modeling are primarily personal health devices used in home and mobile care contexts. 
BlockIoT based data template representation opens doors for new applications that can support intelligent analytics, decision support, machine learning to enable precision medicine.

However, we must note that there are also considerable barriers to getting this system off the ground.
Appropriate security and privacy to comply with health information laws are a must for any application in this area.
As healthcare is a heavily regulated industry, smart contracts operating on a decentralized system that sees no boundaries for data exchange and transaction confirmation would need to consider jurisdictional aspects and institutional policies.
Even if such regulatory issues are resolved, the sheer number of personal health devices and the lack of medical device standards are currently the most significant impediments.

For future work, we plan to extend the data templates to cover various devices beyond the templates noted in \Cref{fig:blockiot-architecture}.
These extensions will require further mappings to semantic web based terminologies that will enhance the expressivity and the interoperability of the templates we will develop.
We also plan to conduct a participatory design evaluation with healthcare providers to ascertain the types of alerts that they prefer to receive based on any analysis of the data collected through the decentralized BlockIoT API. 
Based on that investigation, we will significantly expand our suite of smart contracts.

\section{Conclusion}

In this paper, we have proposed a system called BlockIoT that bridges the gap between personal health data held by the patients and the healthcare providers' electronic health records.
Current information systems and health standards cannot exploit the data, primarily available through wearable devices, efficiently. 
The main reason is that these systems cannot interpret the semantic content of the data, and also the structure of the data and the metadata that describe them are not structured in a formal and unified way. 
These drawbacks can slow down diagnostic procedures, especially for patients suffering from chronic diseases. 
The use of semantic web technologies can create the infrastructure for homogeneous information access and enhance systems interoperability.
Knowledge-driven systems for decision-making in health care applications are powerful tools to help provide actionable and explainable insights to patients and practitioners. 
In such systems, knowledge about the particular patient that includes current condition, historical ailments, etc., is central to enable personalized health care.

The combination of blockchain and semantic web technologies can be considered so vital that it would mark a transformation to a patient-centric health system allowing individuals to control the rights and access to their medical data. 
In aggregate, there are diverse and exceptional opportunities to improve efficiency, accuracy, and workflow in healthcare, improve patient outcomes and lower costs.
BlockIoT is designed to provide real-time medical device data and insights powered by smart contracts to a healthcare provider, who are almost always constrained by their hectic schedules, in an easy-to-understand manner to use this data to make better decisions enabling high-quality and efficient care.

We believe that the systematic application of advanced expressive knowledge management approaches, coupled with a trustworthy blockchain architecture that supports extensible features via smart contracts, can lead to knowledge-mediated patient care planning systems.
These would automatically and proactively generate adaptive, personalized healthcare suggestions that may guide the long-term clinical, therapeutic care process for individual patients within a specific healthcare setting.

The open-source software of the BlockIoT system described in this paper is available at \url{https://github.com/rpi-scales/BlockIoT}.
\balance
\bibliographystyle{ACM-Reference-Format}
\bibliography{references.bib}

\end{document}